\documentclass[slac_one]{revtex4}
\usepackage{graphicx}
\usepackage{fancyhdr}
\usepackage[centertags]{amsmath}
\usepackage{amsfonts}
\usepackage{amssymb}
\usepackage{amsthm}
\begin{document}

\title{Electromagnetically Interacting Massive Spin-2 Field: Intrinsic
Cutoff and Pathologies in External Fields} 

%

\author{Massimo Porrati, Rakibur Rahman}
\affiliation{ CCPP, Department of Physics, New York University, New York, NY 10003, USA}

\begin{abstract}
By employing the St\"uckelberg formalism, we argue that the theory of massive
spin-2 field coupled to electromagnetism in flat space must have an intrinsic,
model independent, finite UV cutoff. We show how the very existence of a cutoff
has connection to other pathologies of the system, such as superluminal propagation.
We comment on the generalization of the results to arbitrary spin, and to
gravitational interaction.
\end{abstract}

\maketitle

\thispagestyle{fancy}


\section{INTRODUCTION} 

Massive charged high-spin particles do exist in nature. Particle colliders have produced
unstable composite high-spin resonances like $\pi_2$(1670), $\rho_3$(1690) or $a_4$(2040),
whose inverse size is of the same order as their mass. String theory also gives massive high-spin
particles, with masses at least as large as the string scale, that may very well couple to
a $U(1)$ gauge field. In both cases the electrodynamics can be described by an effective field
theory, which makes sense only below a finite cutoff not higher than the mass scale itself.
One may wonder whether this is a generic feature of massive charged high-spin fields, or just
particular to the above two examples.

On the other hand, massive charged fields with $s>1$ are known to exhibit a variety of pathological
features in a constant electromagnetic background in flat space~\cite{vz}. The high-spin field may
suffer from the so-called Velo-Zwanziger acausality~\cite{vz} with modes that propagate superluminally,
or the system even ceases to be hyperbolic, or the number of propagating degrees of freedom may be
different from that of the free theory, or the Cauchy problem may become ill-posed. One would like
to know if the possible existence of a cutoff is related to all these pathologies.

The coupling of \emph{massless} high-spin particles to electromagnetism or gravity is
plagued by grave inconsistencies~\cite{ad}, because of absence of conserved current invariant under
the high-spin gauge symmetry. For gravitational and electromagnetic interactions this problem
respectively starts from spin-5/2~\cite{ad}, and spin-3/2 (in flat space). In fact, there are strong
constraints coming from no-go theorems~\cite{ww,p} that forbid the existence of interacting massless
particles with $s>2$ in Minkowski space. The fact that charged high-spin resonances exist then
implies that any local Lagrangian describing electromagnetically interacting massive high-spin fields
must be singular in the $m\rightarrow 0$ limit.

In the unitary gauge this massless singularity is not obvious at all. For example, if one
starts with the Singh-Hagen Lagrangian~\cite{sh} for massive fields of arbitrary spin and
introduces electromagnetic coupling by the minimal prescription $\partial_\mu\rightarrow D_\mu$,
one only gets positive powers of the mass in the resulting Lagrangian. Propagators of the
massive theory become singular in the massless limit because of high-spin gauge invariance,
so that scattering amplitudes diverge. The best way to understand the mass singularity
is to employ the St\"uckelberg formalism, because it focuses precisely on the gauge modes
responsible for bad high energy behavior. Here one renders the massive free theory gauge
invariant through the addition of auxiliary fields, which can always be set to vanish using
the resulting gauge invariance. When the theory is coupled to a $U(1)$  gauge field, one can
make a judicious covariant gauge fixing to obtain diagonal kinetic operators.
In this case, there exist non-renormalizable interaction terms involving auxiliary fields,
which explicitly depend on inverse powers of the mass.

In this article we consider the simple but nontrivial case of
massive spin-2 field coupled to electromagnetism in flat space. In
Section 2 we describe massive spin-2 field \`{a} la St\"uckelberg,
and follow the prescription outlined in~\cite{pr1} to show that the
massless singularity cannot be cured by any means. We quantify the
essential degree of this singularity by finding an expression for
the UV cutoff of the effective action in terms of the particle's
mass and electric charge. In Section 3 we argue that, for a
judicious choice of non-minimal interaction terms, the system may
show pathologies only when one extrapolates the local effective
description beyond its regime of validity. We draw concluding
remarks in Section 4, and comment on the generalization of the
results to arbitrary spin and to gravitational coupling.

\section{ELECTROMAGNETIC COUPLING OF MASSIVE SPIN-2 FIELD}

Massive spin-2 electromagnetic interaction has been studied by various authors
~\cite{vz,pr1,dw,fksc,z}. Here we will always work in flat space. We start with the Pauli-Fierz
Lagrangian~\cite{pf} for massive spin-2 field.
\begin{eqnarray}\label{r1} L= -\frac{1}{2}(\partial_\mu h_{\nu\rho})^2+(\partial_\mu
h^{\mu\nu})^2 +\frac{1}{2}(\partial_\mu h)^2\,-\partial_\mu h^{\mu\nu}\partial_\nu h
-\frac{m^2}{2}[h_{\mu\nu}^2-h^2],\end{eqnarray} where $h=h^\mu_{~\mu}$. This Lagrangian
does not possess any manifest gauge invariance. Now, by the field redefinition
\begin{eqnarray}\label{r2} h_{\mu\nu}\rightarrow\tilde{h}_{\mu\nu}=h_{\mu\nu} +\frac{1}{m}\,
\partial_{\mu}\left(B_\nu-\frac{1}{2m}\partial_{\nu}\phi\right)+\frac{1}{m}\,\partial_{\nu}
\left(B_\mu-\frac{1}{2m}\partial_{\mu}\phi\right),\end{eqnarray} we introduce new (St\"{u}ckelberg)
fields $B_{\mu}$ and $\phi$ to obtain the following gauge invariance (St\"{u}ckelberg symmetry): \begin{eqnarray}\label{r3}\delta h_{\mu\nu}&=&\partial_{\mu}\lambda_{\nu}+\partial_{\nu}
\lambda_{\mu},\\\label{r4}\delta B_{\mu} &=& \partial_{\mu}\lambda - m\lambda_{\mu},\\\label{r5}
\delta\phi &=& 2m\lambda.\end{eqnarray} The St\"{u}ckelberg fields are redundant, in that they can
always be gauged away. Yet they are useful since they allow us to unveil the dangerous degrees of
freedom and interactions hidden inside the spin-2 action. In Lagrangian~(\ref{r1}) all the higher
dimensional operators cancel by the commutativity of partial derivatives. In fact one can
obtain the St\"{u}ckelberg version of the Pauli-Fierz Lagrangian just by Kaluza-Klein reducing the
(4+1)D linearized Einstein-Hilbert action to (3+1)D~\cite{adyrs}. All the fields $h_{\mu\nu}$,
$B_{\mu}$ and $\phi$ then originate from a single higher dimensional massless spin-2 field, and the
higher dimensional gauge invariance gives rise to the St\"{u}ckelberg symmetry~(\ref{r3}-\ref{r5})
in lower dimension.

Electromagnetic coupling is introduced by complexifying the fields and replacing ordinary derivatives
with covariant ones.
\begin{equation}\label{r6} L= -\,|D_\mu\tilde{h}_{\nu\rho}|^2+2|D_\mu \tilde{h}^{\mu\nu}|^2
+|D_\mu\tilde{h}|^2 - [D_\mu \tilde{h}^{*\mu\nu}D_\nu \tilde{h} + \text{c.c.}] -m^2[\tilde{h}_{\mu\nu}^*
\tilde{h}^{\mu\nu}-\tilde{h}^*\tilde{h}]-\frac{1}{4}\,F_{\mu\nu}^2,\end{equation} where $\tilde{h}_{\mu\nu}$, the covariant counterpart
of~(\ref{r2}),
\begin{equation}\label{r7} \tilde{h}_{\mu\nu} =h_{\mu\nu}+
\frac{1}{m}\,D_{\mu}\left(B_\nu-\frac{1}{2m}D_{\nu}\phi\right) +\frac{1}{m}\,D_{\nu}\left(B_\mu-\frac{1}
{2m}D_{\mu}\phi\right),\end{equation} is trivially invariant under the gauged version of the
symmetry~(\ref{r3}-\ref{r5}). The authors in Ref.~\cite{z} also considered a gauge invariant description
to investigate consistent interactions of massive high-spin fields. In case of spin-2, say, they introduce St\"{u}ckelberg fields only in the mass term. This procedure already breaks St\"{u}ckelberg invariance at
tree level, but our approach by construction guarantees that St\"{u}ckelberg symmetry is intact by the
minimal substitution.

Next, we diagonalize the kinetic operators to make sure that the propagators in the theory have good high
energy behavior, i.e., that all propagators are proportional to $1/p^2$ for momenta $p^2\gg m^2$. The field
redefinition $h_{\mu\nu}\rightarrow h_{\mu\nu}-(1/2)\eta_{\mu\nu}\phi$ eliminates some kinetic
mixings, and also generates a kinetic term for $\phi$ with the correct sign. Adding the gauge fixing terms: $L_{\text{gf1}}=-2|\partial_\nu h^{\mu\nu}-(1/2)\partial^\mu h+mB^\mu|^2$, and $L_{\text{gf2}}=-2\,|\partial_\mu B^{\mu}+(m/2)(h - 3\phi)|^2$, we exhaust all gauge freedom to obtain diagonal kinetic terms. We are left with \begin{eqnarray}\label{r8} L=h_{\mu\nu}^{*}(\Box-m^2)h^{\mu\nu}-\frac{1}{2}h^*(\Box-m^2)h + 2B_\mu^*(\Box-m^2)
B^\mu+\frac{3}{2}\phi^{*}(\Box-m^2)\phi-\frac{1}{4}F_{\mu\nu}^2 + L_{\text{int}}.\end{eqnarray}
Here $L_{\text{int}}$ contains all interaction terms, among which there exist higher dimensional operators,
thanks to the non-commutativity of covariant derivatives. The most potentially dangerous non-renormalizable
operators in the high energy limit $m\rightarrow0$ are the ones with the highest dimensionality. Parametrically
in $e\ll 1$, for any given operator dimensionality, the $\mathcal{O}(e)$-terms are more dangerous than the
others. We have \begin{eqnarray}\label{r9} L_{\text{int}}=\frac{e}{m^4}\,\partial_\mu F^{\mu\nu}[(i/2)
\partial_\rho\phi^*\partial^\rho\partial_\nu\phi+\text{c.c.}]+\frac{e}{m^3}F^{\mu\nu}[i(2\partial_\mu B^*_\rho\partial^\rho\partial_\nu\phi-\partial_\mu B_\nu^*\Box\phi)+\text{c.c.}]+~(...),\end{eqnarray}
where (...) stands for less divergent terms. The $\mathcal{O}(e)$ dimension-8 operator is proportional the
Maxwell equations; it can be eliminated by a local field redefinition: $A_\mu\rightarrow A_\mu-(e/m^4)J_\mu$,
where $J_\mu\equiv[(i/2)\partial_\rho\phi^*\partial^\rho\partial_\mu\phi+\text{c.c.}]$. This introduces
$\mathcal{O}(e^2)$-terms: $(e^2/4m^8)[\partial_\mu J_\nu-\partial_\nu J_\mu]^2$, which must be canceled as
well. Indeed, adding the local function: $L_{\text{add}}=(e^2/4)[\tilde{h}^*_{\mu\rho}\tilde{h}^\rho_{~\nu}
-\tilde{h}^*_{\nu\rho}\tilde{h}^\rho_{~\mu}]^2$ to the Lagrangian serves the purpose. But now we will have
terms proportional to $e^2/m^7$, coming both from $L_{\text{add}}$, and from the shift of $A_\mu$
acting on the dimension-7 operator in~(\ref{r9}). The latter gives
\begin{eqnarray}\label{r10} L_{11}=\frac{e^2}{m^7}\,\{\partial^\mu\partial_\sigma\phi^*\partial^\nu\partial^\sigma\phi
-(\mu\leftrightarrow\nu)\}\,\{2\partial_\mu B^*_\rho\partial^\rho\partial_\nu\phi\ - \partial_\mu B_\nu^*\Box \phi\}+\text{c.c.} \end{eqnarray} Notice that any \emph{local} function of the spin-2 field $\tilde{h}_{\mu\nu}$
is fully invariant under the special St\"uckelberg symmetry:
\begin{eqnarray}\label{r11} B_\mu\rightarrow B_\mu+b_\mu+(b_{\mu\nu}-b_{\nu\mu})x^\nu,~~~~\phi\rightarrow \phi
+c+c_\mu x^\mu,\end{eqnarray} where $b_\mu, b_{\mu\nu}$, $c$, and $c_\mu$ are constant tensors. Now the
operators in Eq.~(\ref{r10}), being invariant only up to a nontrivial total derivative, cannot be eliminated
by adding local functions of $\tilde{h}_{\mu\nu}$. We must have terms proportional to $e^2/m^7$.

On the other hand, mere addition of a dipole term leaves us only
with terms proportional to $e/m^3$, which is already an improvement
over field redefinition plus addition of local term. Indeed, a
dipole term $2ie\alpha
F^{\mu\nu}\tilde{h}^*_{\mu\rho}\tilde{h}^\rho_{\;\;\nu}$ gives
\begin{eqnarray}\label{r12}
L_{\text{dipole}}=\frac{e}{m^4}\,\partial_\mu
F^{\mu\nu}[-i\alpha\partial_\rho\phi^*\partial^\rho
\partial_\nu\phi+\text{c.c.}]+\frac{e}{m^3}\,F^{\mu \nu}[-2i\alpha\partial_{(\mu}B^*_{\rho)}\partial^\rho
\partial_\nu\phi+\text{c.c.}]+(...)\end{eqnarray} If we choose $\alpha=1/2$,
the dimension-8 operators at $\mathcal{O}(e)$ all cancel leaving us only with dimension-7 operators.
In the scaling limit: $m\rightarrow0$, $e\rightarrow0$, such that $e/m^3$=constant, the non-minimal
Lagrangian reduces to: \begin{eqnarray}\label{r13}
L=L_{\text{kin}}+\frac{ie}{2m^3}\,F^{\mu\nu}\left\{2\partial_{[\mu}B_{\rho]}^*\partial^\rho\partial_\nu\phi
-\partial_{[\mu}B_{\nu]}^*\Box\phi\right\}+ \text{c.c.}\end{eqnarray} These operators contain pieces that
are not proportional to the equations of motion. Therefore the degree of divergence could not have been
improved further. Thus our theory has an intrinsic model-independent cutoff~\cite{opt}:
\begin{eqnarray}\label{r14} \Lambda=\frac{m}{e^{1/3}}~.\end{eqnarray} Here we notice the
curious fact that the Lagrangian (\ref{r13}) has acquired a $U(1)$ gauge invariance for the vector
St\"{u}ckelberg $B_\mu$. This is because an appropriately chosen dipole term cancels not only the
$\mathcal{O}(e)$ dimension-8 operator, but also any $\mathcal{O}(e)$ dimension-7 operators that get
generated from a transformation $B_{\mu}\rightarrow B_{\mu}+\partial_{\mu}\theta$.

\section{SUPERLUMINAL PROPAGATION AND ALL THAT}

All the pathologies of interacting high-spin system originate from a
gauge invariance in the kinetic part of the free Lagrangian, which
implies the existence of zero modes. In electromagnetic backgrounds
the zero modes acquire non-vanishing but non-canonical kinetic term,
which may cause some of them to propagate faster than light, or even
lose hyperbolicity. The St\"uckelberg formalism is tailored to
single out the dynamics of precisely these gauge modes, which are
nothing but the St\"uckelberg fields. For the case of massive
spin-2, to simplify our analysis we set the $B_\mu$ field on shell.
By doing this we can only check if the scalar $\phi$ exhibits
non-standard dynamics. In a \emph{constant} electromagnetic
background $\phi$ will experience a new effective background metric,
different from Minkowski, because of some $\mathcal{O}(e^2)$
dimension-8 operators~\cite{pr1}:
\begin{eqnarray}\label{r15}\tilde{\eta}^{\mu\nu}=
\left\{\frac{3}{2}+\frac{e^2}{4m^4}(5-4\alpha)
F_{\rho\sigma}^2\right\}\eta^{\mu\nu}+\frac{e^2}{4m^4}(2+8\alpha)F^{\mu\rho}F_\rho^{\;\;\nu},\end{eqnarray}
where $\alpha$ is a generic dipole coefficient.
$\tilde{\eta}^{\mu\nu}$ is proportional to the Minkowski metric for
$\alpha=-1/4$, so that $\phi$ propagates at the speed of light. But
even for this value of $\alpha$ the system ceases to be hyperbolic
in a strong field; precisely when $e^2 F_{\mu\nu}^2/m^4=-1$. By
computing the characteristic speeds one finds that the choice
$\alpha<-1/4$ is downright pathological~\cite{pr1}, because it gives
superluminal propagation even for infinitesimally small values of
the background fields. Other values of $\alpha$ are safe for
sufficiently weak external fields, and pathologies exist only in
strong external fields: $E^2,B^2\gtrsim (m/\sqrt{e})^4>\Lambda^4$.
But this is already beyond the regime of validity of our effective
field theory. In that dynamical regime, pathologies may be cured by
new degrees of freedom or strong coupling phenomena.

\section{CONCLUSION}

The St\"uckelberg method is powerful in that it clarifies the
dynamics of interacting high-spin fields. Here we analyzed a simple
high-spin system: a spin-2 field coupled to electromagnetism and
found that the theory must have an intrinsic UV cutoff, no higher
than $\Lambda=me^{-1/3}$. The pathologies manifest themselves in the
scalar St\"uckelberg sector. However we saw that, for a judicious
choice of non-minimal interaction terms, they appear in a regime
where we can no longer trust our effective field theory description,
at least in the scalar sector.

Generalizations of this example to arbitrary spin $s$ is also being done~\cite{pr2}. Starting from spin-5/2
the story gets complicated by the presence of auxiliary fields that are not (gamma)traces of the high-spin
field. The cutoff turns out to be no higher than $\Lambda=me^{-1/(2s-1)}$. This result rules out long-lived
high-spin charged particles, and thus may affect directly the strategy for their search in future colliders.
Gravitational coupling of high-spin fields is also very important; it should be relatively straightforward
in our formalism. Existence of an intrinsic cutoff in this case will signal the ultimate limit of any local
effective field theory description of interacting massive high-spin fields.

Does the existence of a UV cutoff necessarily call for new physics? Suppose some nontrivial UV fixed point
exists~\cite{uv}. Then it means that the theory can be resummed, so that the massless limit will be smooth.
While this might be true for a spin-2 field with non-minimal electromagnetic interaction, absence of smooth
massless limit starting from spin-5/2~\cite{ad,p} rules out such a possibility for higher-spins.

\begin{acknowledgments}
R.R. wishes to thank the ICHEP'08 organizers for their kind hospitality.
M.P. would like to thank the Galileo Galilei Institute for Theoretical Physics for hospitality and INFN for
partial support during the completion of this work. M.P. is supported in part by NSF grants PHY-0245068 and PHY-0758032.
\end{acknowledgments}


\begin{thebibliography}{99}   
\bibitem{vz}
  G.~Velo and D.~Zwanziger,
  Phys.\ Rev.\  {\bf 186}, 1337 (1969);
  Phys.\ Rev.\  {\bf 188}, 2218 (1969).
  G.~Velo,
  Nucl.\ Phys.\  B {\bf 43}, 389 (1972).
\bibitem{ad}
  C.~Aragone and S.~Deser,
  Phys.\ Lett.\  B {\bf 86}, 161 (1979).
\bibitem{ww}
  S.~Weinberg and E.~Witten,
  Phys.\ Lett.\  B {\bf 96}, 59 (1980).
\bibitem{p}
  M.~Porrati,
  arXiv:0804.4672 [hep-th].
\bibitem{sh}
  L.~P.~S.~Singh and C.~R.~Hagen,
  Phys.\ Rev.\  D {\bf 9}, 898 (1974);
  Phys.\ Rev.\  D {\bf 9}, 910 (1974).
\bibitem{pr1}
  M.~Porrati and R.~Rahman,
  Nucl.\ Phys.\  B {\bf 801}, 174 (2008)
  [arXiv:0801.2581 [hep-th]].
\bibitem{dw}
  S.~Deser and A.~Waldron,
  Nucl.\ Phys.\  B {\bf 631}, 369 (2002)
  [arXiv:hep-th/0112182];
  Phys.\ Rev.\  D {\bf 74}, 084036 (2006)
  [arXiv:hep-th/0609113].
\bibitem{fksc}
  P. Federbush,
  Nuovo Cimento {\bf 19}, 572 (1961).
  M.~Kobayashi and A.~Shamaly,
  Phys.\ Rev.\  D {\bf 17}, 2179 (1978);
  Prog.\ Theor.\ Phys.\  {\bf 61}, 656 (1979).
  A.~Shamaly and A.~Z.~Capri,
  Annals Phys.\  {\bf 74}, 503 (1972).
\bibitem{z}
  S.~M.~Klishevich and Yu.~M.~Zinovev,
  Phys.\ Atom.\ Nucl.\  {\bf 61}, 1527 (1998)
  [Yad.\ Fiz.\  {\bf 61}, 1638 (1998)]
  [arXiv:hep-th/9708150];
  Yu.~M.~Zinoviev,
  Nucl.\ Phys.\  B {\bf 770}, 83 (2007)
  [arXiv:hep-th/0609170].
  Yu.~M.~Zinoviev,
  arXiv:0806.4030 [hep-th].
\bibitem{pf}
  M.~Fierz and W.~Pauli,
  Proc.\ Roy.\ Soc.\ Lond.\  A {\bf 173}, 211 (1939);
  Helv.\ Phys.\ Acta {\bf 12}, 297 (1939).
\bibitem{adyrs}
  C.~Aragone, S.~Deser and Z.~Yang,
  Annals Phys.\  {\bf 179}, 76 (1987).
  S.~D.~Rindani and M.~Sivakumar,
  Phys.\ Rev.\  D {\bf 32}, 3238 (1985);
  S.~D.~Rindani, D.~Sahdev and M.~Sivakumar,
  Mod.\ Phys.\ Lett.\  A {\bf 4}, 265 (1989).
\bibitem{opt}
  This is the true cutoff of the theory, while the ``optimistic" one: $\Lambda_2=m/\sqrt{e}$,
  mentioned in \cite{pr1} is too optimistic.
\bibitem{pr2}
  M. Porrati and R. Rahman, work in progress.
\bibitem{uv}
  We would like to thank J. Polchinski for pointing out to us that the strong coupling dynamics
  of our theory may generate a UV fixed point.

\end{thebibliography}
\end{document}